\documentclass[prl,twocolumn,showpacs,superscriptaddress,preprintnumbers,nofootinbib, amsmath,amssymb]{revtex4-2}

\usepackage[utf8]{inputenc}
\usepackage{color}
\usepackage{amsmath}
\usepackage{amssymb}
\usepackage{graphicx}
\usepackage{braket}
\usepackage{tikz}
\usepackage{bm}
\usepackage[unicode=true,pdfusetitle,
 bookmarks=true,bookmarksnumbered=false,bookmarksopen=false,
 breaklinks=false,pdfborder={0 0 0},pdfborderstyle={},backref=false,colorlinks=true]
 {hyperref}
\hypersetup{
 pdfborderstyle={},colorlinks,linkcolor=red,citecolor=blue}

\makeatletter

\usepackage{setspace}
\usepackage{dcolumn}
\usepackage{mathrsfs}
\usepackage{siunitx}
\makeatother

\begin{document}
\title{Dark Matter Detection through Rydberg Atom Transducer}

% Replace the existing authorship block with the following

\author{J. F. Chen}
\affiliation{Department of Physics, Southern University of Science and Technology, Shenzhen 518055, China}
\affiliation{International Shenzhen Quantum Academy, Shenzhen 518045, China}

\author{Haokun Fu}
\affiliation{Department of Physics, Southern University of Science and Technology, Shenzhen 518055, China}

\author{Christina Gao}
\email{gaoy3@sustech.edu.cn}
\affiliation{Department of Physics, Southern University of Science and Technology, Shenzhen 518055, China}

\author{Jing Shu}
\affiliation{School of Physics and State Key Laboratory of Nuclear Physics and Technology, Peking University, Beijing 100871, China}
\affiliation{Center for High Energy Physics, Peking University, Beijing 100871, China}
\affiliation{Beijing Laser Acceleration Innovation Center, Huairou, Beijing 101400, China}

\author{Geng-Bo Wu}
\affiliation{State Key Laboratory of Terahertz and Millimeter Waves, City University of Hong Kong, Hong Kong 999077, China}
\affiliation{Department of Electrical Engineering, City University of Hong Kong, Kowloon, Hong Kong SAR, China}
\affiliation{City University of Hong Kong Shenzhen Research Institute, Shenzhen 518057, China}

\author{Peiran Yin}
\affiliation{National Laboratory of Solid State Microstructures and Department of Physics, Nanjing University, Nanjing, China}

\author{Yi-Ming Zhong}
\affiliation{Department of Physics, City University of Hong Kong, Kowloon, Hong Kong SAR, China}

\author{Ying Zuo}
\email{zuoying@iqasz.cn}
\affiliation{International Shenzhen Quantum Academy, Shenzhen 518045, China}

\date{\today}

% ==============================================================================
% Abstract
% ==============================================================================
\begin{abstract}
Ultralight bosonic dark matter with masses in the meV range, corresponding to terahertz (THz) Compton frequencies, remains largely unexplored due to the difficulty of achieving both efficient signal conversion and single-photon-sensitive detection at THz frequencies. We propose a hybrid detection architecture that integrates a dielectric haloscope, Rydberg-atom transducer, and superconducting nanowire single-photon detection within a unified cryogenic platform operating at $\lesssim 1\,\text{K}$. The dielectric haloscope converts dark matter into THz photons via phase-matched resonant enhancement, achieving form factors $C \sim 0.4$ and loaded quality factors $Q_L \sim 10^4$. A cold $^{87}$Rb ensemble then coherently up-converts the THz signal to the optical domain through six-wave mixing among Rydberg states. The intrinsic directionality and narrow bandwidth ($\Delta\nu_{\mathrm{atomic}} \sim 1\,\text{MHz}$) of this process provide extra suppression of isotropic thermal backgrounds. 
%We identify 798 dipole-allowed Rydberg transitions spanning $0.1-1.5\,\text{THz}$, enabling broadband mass coverage by stepping through discrete Rydberg state pairs. 
With 10 days of integration at $0.3$~K, we project sensitivity to the axion-photon coupling $g_{a\gamma\gamma} \sim 10^{-13}\,\mathrm{GeV}^{-1}$ at $m_a \sim 0.4\,\text{meV}$, reaching the QCD axion band and opening the THz window for searches of both axion and dark photon dark matter.
\end{abstract}

\maketitle

% ==============================================================================
% Introduction — 3 paragraphs
% ==============================================================================

Dark matter (DM) constitutes about 85\% of matter in our Universe, yet its nature remains elusive~\cite{Bertone:2016nmp}. Among the compelling candidates are ultralight bosons, including axions~\cite{Kim:1979if,Shifman:1979if,Zhitnitsky:1980tq,Dine:1981rt}, axion-like particles (ALPs)~\cite{Witten:1984dg, Svrcek:2006yi, Arvanitaki:2009fg}, and dark photons~\cite{Abel:2008ai,Nelson:2011sf,Arias:2012az}, all arising naturally in extensions of the Standard Model. 
 Ultralight DM shares the feature that, at sub-eV masses, they exhibit wave-like behavior: the DM field oscillates coherently at a frequency set by its mass.
This wave-like nature enables a distinctive search strategy compared to weakly-interacting massive particle (WIMP) searches. When coupled to photons, ultralight DM acts as an effective current source in Maxwell's equations, capable of driving detectable electromagnetic (EM) fields in laboratory resonators tuned to the DM mass~\cite{PhysRevLett.51.1415,Graham:2014sha,Arias:2014dla}. 
Using microwave cavity-based haloscopes, substantial sensitivities to axion-photon coupling, $g_{a\gamma\gamma}$, and dark photon-photon kinetic mixing, $\chi$, have been achieved for DM masses around $\mu$eV~\cite{PhysRevLett.120.151301,PhysRevLett.124.101303,CAPP:2024axion,HAYSTAC:2021squeezing,HAYSTAC:2024phaseII,ORGAN:2024,QUAX:2024,MADMAX:2024first,CAST-RADES:2021,DarkSRF:2023,SHANHE:2024}. Meanwhile, telescope and helioscope searches have probed ultralight DM  with masses above $\sim$100~meV~\cite{CAST:2014buf,Janish:2023kvi,Pinetti:2025owq}.

\begin{figure*}[t]
    \centering
    \includegraphics[width=0.5\textwidth]{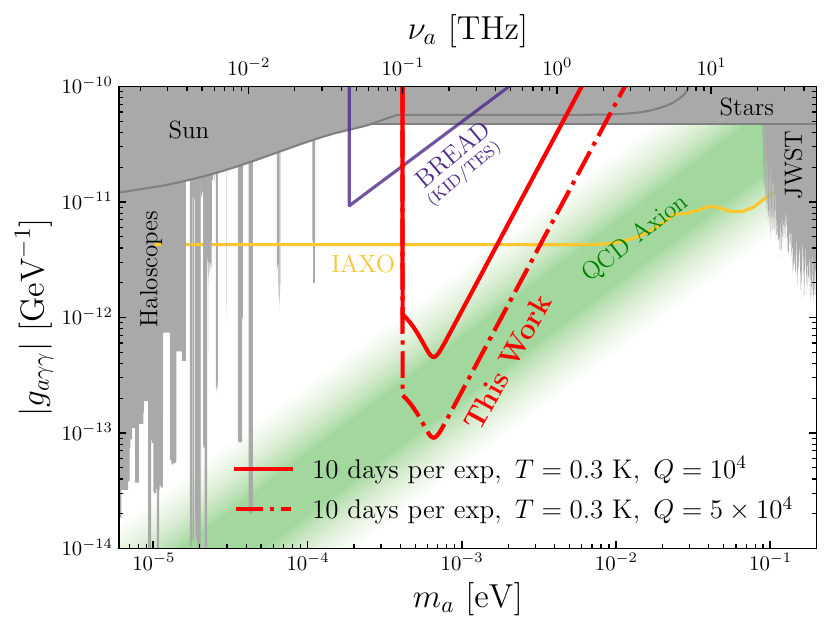}~\includegraphics[width=0.5\textwidth]{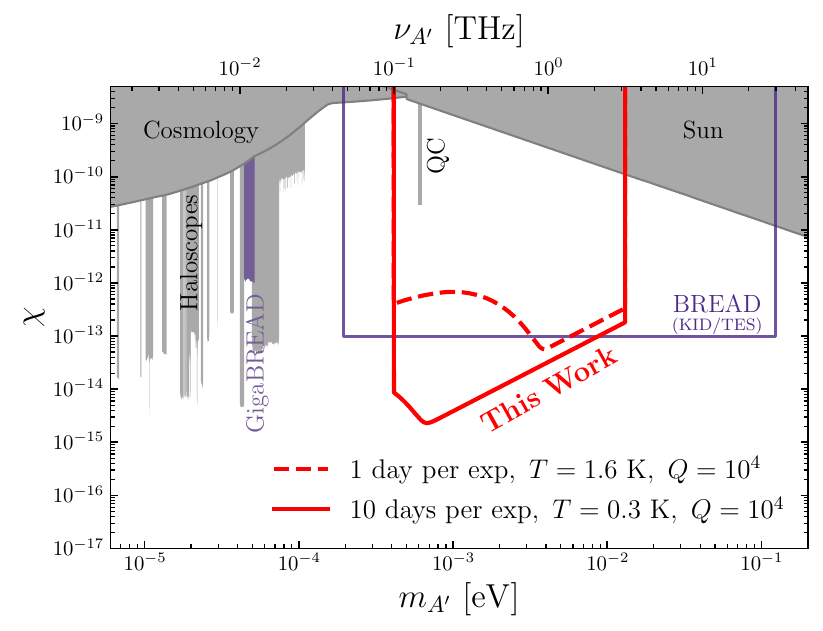}
    \caption{Projected sensitivities of our detection setup via DM-THz-Optical conversion for axion DM, $a$, (\emph{left}) and dark photon DM, $A'$(\emph{right}), assuming $C=0.4$ and five layers for the dielectric haloscope. For the axion DM search, we further assume that $B_0=10$~T,  an integration time of $t=10$ days per experiment, and $T = 0.3\,\text{K}$ with $Q=10^4$ (solid) or $Q=5\times 10^4$ (dash-dotted). For the dark photon DM search, we assume $t=1$ day per experiment, and $Q=10^4$  
    at $T = 1.6$ K (dashed) or $T=0.3$ K (solid). %Approximately $10^5$ experimental setups are needed for each curve. 
    Gray shading shows existing constraints from~\cite{AxionLimits}. The green band and yellow solid lines in the left panel show the predictions of QCD axion models~\cite{DiLuzio:2016sbl} and the projected sensitivity of IAXO~\cite{IAXO:2019mpb}, respectively. The purple shading in the right panel shows constraints from GigaBREAD~\cite{BREAD:2023xhc}.
    The purple lines in both panels show the projected sensitivities of BREAD with KIDS/TES (10 days, 0.3 K)~\cite{BREAD:2021tpx}.}
    \label{fig:sensitivity}
\end{figure*}

However, for DM masses around $0.1$--$10$ meV, which correspond to Compton frequencies ($\nu_\text{DM} = m_\text{DM}/(2\pi)$\footnote{Below we use natural units $c=\hbar =1$.}) in the terahertz (THz) range, the ultralight DM remains largely unexplored.
This gap persists due to two obstacles. 
First, there is a mismatch between the dispersion relations of the DM particles, which are non-relativistic ($v \sim 10^{-3}$~\cite{Evans:2018bqy}), and free photons. This causes the DM-photon conversion efficiency to be strongly suppressed, unless the photon's dispersion relation is significantly altered, as in the case of traditional cavity-based searches where the size of a cavity is on the order of DM Compton wavelength $2\pi /m_\text{DM}$. 
But at THz frequencies, such a cavity would imply a $\mathrm{mm}^3$ volume, thus severely limiting the signal power. To overcome this obstacle, several proposals employ dielectric haloscopes~\cite{2017DielectricHaloscopesNewWayDetectAxionDarkMatter,Millar:2016cjp,Baryakhtar:2018dxp,Brun:2019lyf,LAMPOST:2022,sdg3-8dpw}, 
which enable phase matching through periodic refractive-index modulation.

Second, unlike the microwave regime where quantum-limited amplifiers are mature~\cite{RevModPhys.82.1155}, single-photon-sensitive detection at THz frequencies remains difficult: bolometric detectors suffer from thermal noise at THz frequencies and quantum-limited amplifiers do not exist in the range~\cite{rogalski_terahertz_2011,Zeuthen_2020}. 
Recent advances in atomic technologies come to the rescue. Rydberg-atom ensembles have emerged as promising platforms for coherent EM field sensing and THz-to-optical transduction~\cite{tu_high-efficiency_2022,kumar_quantum-enabled_2023,borowka_continuous_2024,li_room_2024,PhysRevLett.120.093201,PhysRevA.100.012307}. 
Several proposals are based on those advances: Rydberg-atom-based single-photon detection for axion haloscope readout in  $10$--$50$~GHz range~\cite{PhysRevD.109.032009}, broadband searches using the BREAD dish antenna in 20--2000~GHz range~\cite{banerjee2025rydbergsinglephotondetection}, direct sensing of dark-photon-induced electric fields via optically trapped Rydberg tweezer arrays~\cite{chigusa2025darkmatterdetectionusing},
and cavity-free detection of axion-induced electric dipole transitions~\cite{PhysRevResearch.6.023017}. These proposals differ in whether the Rydberg atoms serve as direct photon absorbers or as coherent transducers, and in whether signal enhancement is from resonant buildup or large geometric collection area, leading to different sensitivity scalings and noise budgets. However, no prior work has integrated a resonant THz conversion stage with an atomic transducer.

In this work, we propose a new architecture that combines a dielectric haloscope, Rydberg-atom transduction, and superconducting nanowire single-photon detection (SNSPD), all operating within a unified cryogenic platform at $\lesssim 1$~K. By up-converting THz photons to the optical range before detection, the scheme circumvents the absence of quantum-limited THz detectors and can achieve a noise floor set by technical dark counts rather than thermal fluctuations for dark matter mass $\gtrsim k_B \times 1\,\text{K}$.
We project sensitivities to the axion-photon coupling $g_{a\gamma\gamma}$ and dark photon-photon kinetic mixing parameter $\chi$ in the THz regime (see Fig.~\ref{fig:sensitivity}), reaching the QCD axion band and probing the largely unexplored THz DM parameter space.

% ==============================================================================
\section*{Axion and Dark Photon Electrodynamics}

The ultralight DM field modifies Maxwell's equations by introducing an effective current density~\cite{PhysRevLett.51.1415,Graham:2014sha,Arias:2014dla}
\begin{equation}
\bm{J}_\mathrm{DM} = \begin{cases}
g_{a\gamma\gamma} \, \bm{B}_0 \, \dot{a}(t) & \text{axion DM}, \\
\chi \, m_{A'}^2 \, \bm{A}'(t) & \text{dark photon DM},
\end{cases}
\label{eq:dm_current}
\end{equation}
where $\bm{B}_0$ is the applied magnetic field and $a~(\bm{A}')$ is the axion (dark photon) field. The amplitude of the field is $\sqrt{2\rho_\mathrm{DM}}/m_{\rm DM}$, where the local DM density $\rho_\mathrm{DM} \approx 0.45~\mathrm{GeV/cm}^3$~\cite{Nesti:2013uwa, Karukes:2019jxv}. When a resonant cavity is tuned to the DM mass $\omega \simeq m_{\rm DM}$, the power of the signal photon $P_T$ can be coherently enhanced:
\begin{equation}
P_T =\rho_\mathrm{DM}VQ_L C\begin{cases}
g_{a\gamma\gamma}^2 B_0^2/m_a ,~\text{axion DM} ;\\
\frac13 \chi^2  m_{A'} , ~\text{dark photon DM},
\end{cases}
\label{eq:axion_power}
\end{equation}
where  $V$ is the cavity volume, $Q_L$ is the cavity's loaded quality factor, and the factor of $1/3$ in the dark photon case arises from averaging over the unknown polarization direction of the dark photon field. $C$ is a form factor quantifying the spatial overlap between the cavity mode and the DM-induced field:
\begin{equation}
    C_{mnp} = \frac{\left|\int d^3x \, \bm{\hat n} \cdot \bm{E}_{mnp}(\bm{x})\right|^2}{ V \int d^3x \, \varepsilon(\bm{x})|\bm{E}_{mnp}(\bm{x})|^2},
    \label{eq:formfactor}
\end{equation}
where $\bm{E}_{mnp}(\bm{x})$ is the electric field of cavity mode $(m,n,p)$, $\hat{\bm{n}}$ is the direction of $\bm{B}_0$ (or the polarization of the dark field), and $\varepsilon(\bm{x})$ is the spatially varying permittivity. 

As discussed above, the momentum mismatch between the uniform DM field and the oscillating cavity mode suppresses $C$ in cavities much larger than $2\pi/m_{\rm DM}$. A dielectric haloscope overcomes the limitation by introducing a periodic refractive-index modulation. Given layer spacing $d$, this provides a reciprocal lattice vector $G \sim 2\pi/d$, which facilitates phase matching and thus allows a coherent signal buildup across a structure much larger than $2\pi/m_{\rm DM}$. 
For benchmark parameters, $V = 3\times 10^{-7}$~m$^3$, $C = 0.4$, $Q_L = 10^4$, $B_0 = 10$~T, at frequency $\nu_\text{DM} \simeq  m_\text{DM}/(2\pi)  = 0.1~\text{THz}$, Eq.~\eqref{eq:axion_power} yields a signal rate of $\sim 10^{-4}$~photons/s at KSVZ sensitivity ($g_{a\gamma\gamma} \sim 10^{-13}$~GeV$^{-1}$). Such a setup therefore demands single-photon-sensitive readout.

% ==============================================================================
% Proposed Setup
% ==============================================================================
\section*{Proposed Setup}

\begin{figure}[t]
    \centering
    \includegraphics[width=1\linewidth]{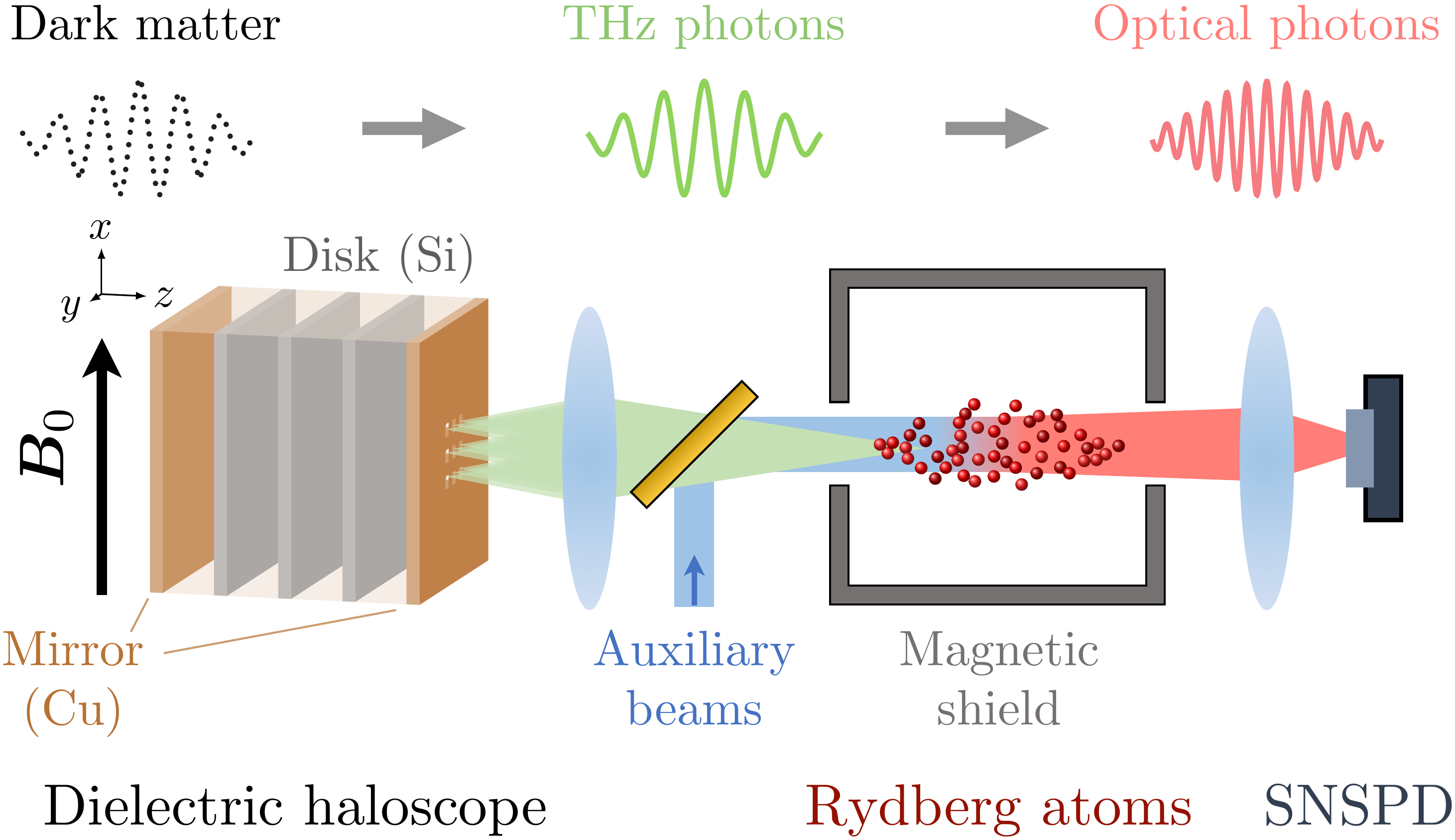}
    \caption{Schematic of the meV-DM cryogenic detection platform. Axion DM  converts to THz photons in a dielectric haloscope, a stack of silicon (Si) disks and copper (Cu) mirrors in a magnetic field $\bm{B}_0$. The THz signal exits through a sub-wavelength $3\times3$ hole array in the end mirror and is focused by a cryogenic lens into a magnetically shielded Rydberg-atom ensemble, where six-wave mixing up-converts it to the optical domain. Spectral filtering and SNSPD readout complete the detection chain. The setup for detecting dark photon DM is similar but no magnetic field is needed for the haloscope.}
    \label{fig:setup}
\end{figure}

The experimental platform, illustrated in Fig.~\ref{fig:setup}, integrates three stages within a unified cryogenic system: a dielectric haloscope for DM-to-THz conversion, a Rydberg-atom ensemble for THz-to-optical transduction, and SNSPDs for low-noise optical readout. For axion DM, the haloscope is immersed in a strong static magnetic field ($B_0 \sim 10$~T from a superconducting solenoid); for dark photon DM, the same apparatus operates in zero magnetic field. We describe each stage below.\\

\begin{figure}[t]
    \centering
    \includegraphics[width=0.8\linewidth]{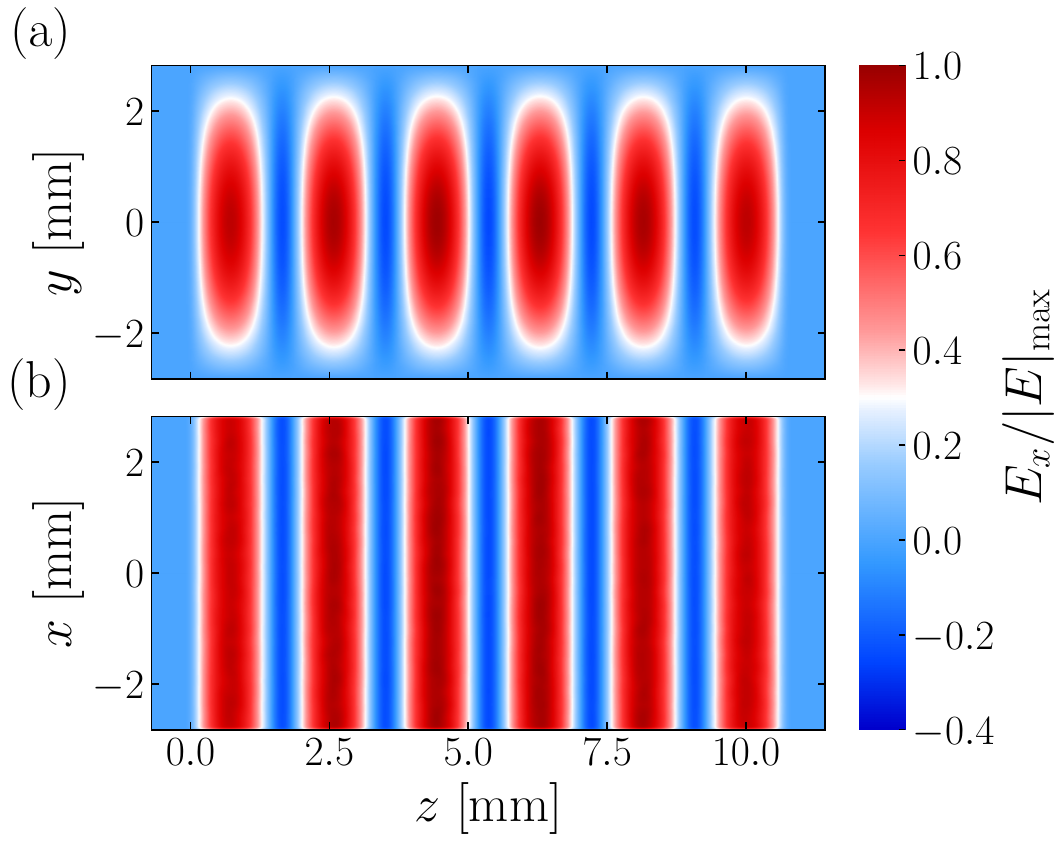}
    \caption{Simulated $E_x$ distribution at $0.1~\mathrm{THz}$, normalized to $|\bm E|_{\max}$. (a) $y$--$z$ cross section at $x=0$. (b) $x$--$z$ cross section at $y=0$. The $E_x$ component, parallel to the applied $\bm{B}_0$, is concentrated in the dielectric layers, yielding $C = 0.37 $.}
    \label{fig:efield}
\end{figure}

\noindent \textit{Dielectric haloscope.} The first stage converts DM into THz photons using a multilayer stack of high-resistivity silicon disks with thickness $d_{\rm Si} \approx \nu_{\rm DM}^{-1}/(2n_\text{Si})$ (refractive index $n_\text{Si}=3.4$) separated by vacuum gaps $d_{\rm vac} \approx \nu_{\rm DM}^{-1}/2$, enclosed by copper mirrors. The cavity has a square base with side length approximately $2\nu_{\rm DM}^{-1}$, and accommodates 5 dielectric layers within a total volume $V \sim \mathcal{O}(10)\nu_{\rm DM}^{-3} $. The cavity design aims to maximize the product $V Q_L C$ while maintaining the resonance frequency within $\pm 30$~MHz of the target frequency. \texttt{COMSOL}~\cite{COMSOL} simulations show that form factors $C \sim 0.4$ and $Q_L \sim 10^4$ are achievable across the $0.1$--$1$ THz range (cf.  App.~\ref{app:cavity}). 
Fig.~\ref{fig:efield} shows an example of the simulated $\bm{E}$ field component that is parallel to the external $\bm{B}_0$. The large asymmetry of the $\bm{E}$ field between the vacuum and silicon layers ensures an order-unity form factor and thus a high DM-to-THz conversion efficiency.
The DM-converted THz photons are then coupled out of the cavity through a $3\times3$ sub-wavelength hole array perforated in the end mirror~\cite{C3689230-232B-4896-9B9E-AF34811493F9}, and focused onto the atomic ensemble by a cryogenic lens~\cite{Scherger:11}.\\

\begin{figure}[t]
    \centering
    \includegraphics[width=1\linewidth]{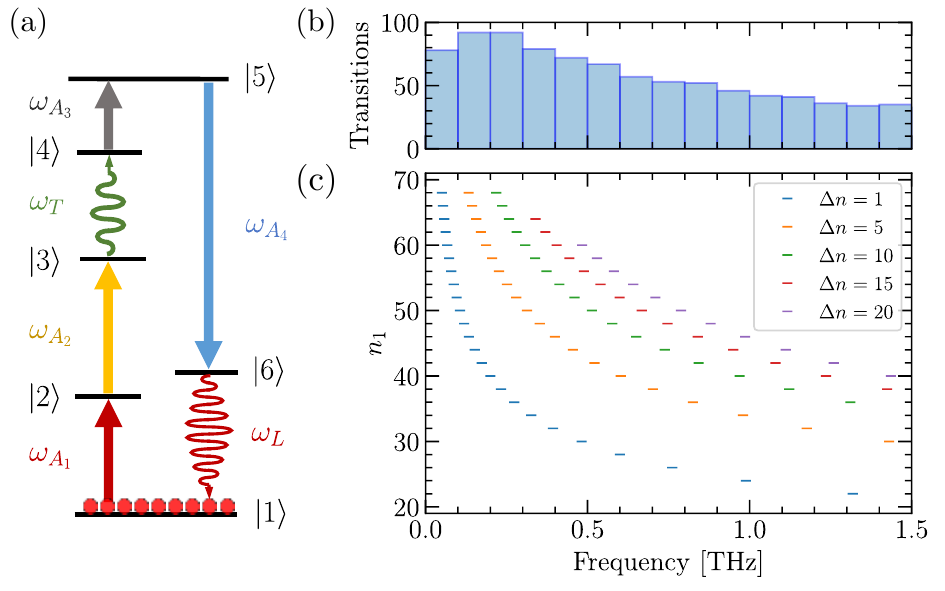}
    \caption{(a) Six-wave-mixing energy-level scheme in $^{87}$Rb, converting THz
photons ($\omega_T=m_{\rm DM}$) to optical photons ($\omega_L$). (b) Transition
frequencies for various Rydberg state combinations. (c) Number of available
transitions per frequency bin; 798 transitions span 0.1--1.5\,THz. All 798
identified transitions are of the $|nS\rangle \to |n'P\rangle$ type.}
    \label{fig:transitions}
\end{figure}

\noindent \textit{Rydberg-atom transducer.} 
A cold $^{87}$Rb ensemble could coherently up-convert THz photons to the optical domain via six-wave mixing (SWM)~\cite{tu_high-efficiency_2022,PhysRevA.110.052608}. 
The atoms are prepared in a cryogenic optical dipole trap or magneto-optical trap and then excited to the Rydberg states with principal quantum number $n$ ranging from 10 to 80~\cite{2010QuantuminformationRydbergatoms}. In the SWM scheme [Fig.~\ref{fig:transitions}(a)], four auxiliary fields ($A_i, i=1,2,3,4$) drive the atoms through six states: $A_1$ and $A_2$ ladder the atom from the ground state $|1\rangle$ up to $|3\rangle$, the incoming THz photon at $\omega_T(=m_{\rm DM})$ couples Rydberg levels $|3\rangle \to |4\rangle$, $A_3$ drives the atom to $|5\rangle$, and $A_4$ brings it to $|6\rangle$, from which it decays back to $|1\rangle$ emitting an optical photon at $\omega_L$. The SWM process offers great spectral flexibility, as the dense manifold of Rydberg states provides a vast range of transitions. 
In particular, the conversion is both directional and narrow-band ($\Delta\nu_{\mathrm{atomic}} \sim 1$~MHz): the phase-matching condition $\bm{k}_{A1} + \bm{k}_{A2} + \bm{k}_T + \bm{k}_{A3} - \bm{k}_{A4} - \bm{k}_L \approx 0$ restricts efficient conversion to a narrow solid angle around the cavity axis, providing additional suppression of isotropic thermal backgrounds.

To estimate the THz-to-optical transduction efficiency, we define
\begin{equation}
    \eta_{\rm trans}= F(L/z_R)\cdot \langle \eta_{\rm int}\rangle\,,
\end{equation}
where $z_R$ is the Rayleigh length for the THz photon, and $L$ is the atomic ensemble length. 
$F(L/z_R)$ is the phase-matching envelope factor~\cite{nonlinearoptics}. At 0.1 THz, $z_R \approx 1.05$~mm, comparable to $L$, so that the Gouy phase shift and wavefront curvature introduce significant spatial mismatch. To suppress this diffraction-induced loss, we employ a compact atomic cloud ($L \lesssim 1$ mm),  and keep the auxiliary beam radii larger than the THz waist to ensure full geometric coverage. In App.~\ref{app:F} we show that $F\geq 66\%$ for frequencies greater than 0.1 THz. $\langle \eta_{\rm int}\rangle\approx\eta_\text{int}^{\max}  \cdot (\Delta \nu_{\mathrm{atomic}}/\Delta \nu_{\mathrm{cav}})$ is the bandwidth-averaged internal efficiency, which takes into account the spectral overlap between the finite atomic transition width $\Delta\nu_{\rm atomic}$ and the cavity photon linewidth $\Delta\nu_{\rm cav}$.
In App.~\ref{app:transduction_eff} we numerically solve the coupled Maxwell-Bloch equations and show that $\eta_\text{int}^{\max}\gtrsim 85\%$ under optimized conditions. 
However, for $Q_L\sim 10^4$, $\Delta\nu_{\rm cav}\gg \Delta\nu_{\rm atomic}$ (cf. Tab.~\ref{tab:atom_params}). This severely degrades the internal efficiency, resulting in $\langle\eta_\text{int}\rangle\gtrsim 1\%$ for signal photons.
Taking both spatial and spectral overlaps into account, we adopt a conservative estimate of transduction efficiency to be
\begin{equation}\label{eq:eta}
    \eta_{\rm trans} \approx 1\% \cdot \left({Q_L}/{10^4}\right),
\end{equation} 
which is valid when $\Delta\nu_{\rm cav}> \Delta\nu_{\rm atomic}$.

To scan the DM mass, the dielectric cavity is retuned in steps of order $\Delta\nu_\mathrm{cav}$, and at each step a Rydberg transition must be available within the cavity linewidth to enable the up-conversion. Since $\Delta\nu_\mathrm{atomic}\ll \Delta\nu_\mathrm{cav}$, the scan rate and mass coverage are set entirely by the cavity tuning; the atomic transitions need only be sufficiently dense that at least one falls within each cavity window. 
Using the \texttt{ARC} package~\cite{SIBALIC2017319}, we compute all dipole-allowed Rydberg transitions with $n_1 = 10$--$70$ and $n_2 = 10$--$80$, identifying 798 transitions within $0.1$--$1.5$ THz as Fig.~\ref{fig:transitions}(b,c) shows.
The transition density is higher at lower frequencies where adjacent Rydberg levels are more closely spaced, and decreases toward 1~THz as larger $\Delta n$ jumps are required. 
Fine-tuning via the Zeeman effect (magnetic fields of $\pm 10$ Gauss, giving $\sim$28 MHz tunability per transition) ensures that at least one atomic resonance can be placed within $\Delta\nu_\mathrm{cav}$ at every target frequency.\\ 

\noindent \textit{Single-photon detection.} The up-converted optical photons are then focused onto the SNSPD, which can achieve detection efficiencies greater than 90\% and intrinsic dark count rates below $10^{-5}$~counts per second (cps) at their sub-kelvin operating temperature~\cite{Zeuthen_2020}.

% ==============================================================================
% Noise and Sensitivity
% ==============================================================================
\section*{Noise Analysis and Sensitivity}

The dominant noise contributions are thermal and technical noises: $\dot{N}_{\text{noise}} = \dot{N}_{\text{thermal}} + \dot{N}_{\text{technical}}$. 
At cryogenic temperatures, thermal photons at THz frequencies are still abundant and can be up-converted by the atomic transducer. Given an effective transduction area $A_\mathrm{eff}$, the thermal noise rate is 
\begin{equation}
\dot{N}_{\text{thermal}}=\frac{2\omega_T^2A_\mathrm{eff}}{\pi (e^{\omega_T/T} -1)}\, \Delta\nu_{\mathrm{atomic}}\,\xi(\omega_T L)\,,
\label{eq:noise_total}
\end{equation}
where 
$\xi$ is a directional suppression factor intrinsic to the SWM process ($\xi\sim 0.1$; see App.~\ref{app:xi} for a full derivation). 

The technical background arises from the stray light leaking from the auxiliary fields $A_{1,2,3,4}$ into the SNSPD, as well as the dark count intrinsic to the SNSPD. 
The stray light can be suppressed via a three-stage filtering scheme combining polarization extinction (50~dB with Glan-Taylor polarizer), laser-line interference filters (60~dB), and high-finesse Fabry-P\'{e}rot cavities (70~dB). Together, these achieve $>$180~dB\footnote{We note that achieving $> 180$~dB total optical isolation in a cryogenic environment has not yet been demonstrated experimentally;
validating this filtering performance is a key technical milestone for the proposed scheme.} of total isolation, suppressing the residual stray-light flux to $ \lesssim 5 \times 10^{-6}$~cps. 
Together with the SNSPD dark count rate $\sim 5 \times 10^{-6}$~cps, the combined technical noise floor is $
\dot{N}_{\text{technical}}\approx 10^{-5}~{\rm cps}$
independent of frequency. \\

\noindent \textit{Projected sensitivity.} The signal-to-noise ratio (SNR) after integration time $t$ is 
\begin{equation}
    \mathrm{SNR} = \frac{\dot{N}_\mathrm{sig}\, t}{
  \mathrm{Max}\!\left[1,\, \dot{N}_\mathrm{noise}\, t\right]^{1/2}}~.
\end{equation}  
The Max function interpolates between the background-free limit, where the sensitivity scales as $\sqrt t$, and the background-dominated regime where the sensitivity scales as $t^{1/4}$. 
The signal rate can be obtained from Eq.~\eqref{eq:axion_power} and Eq.~\eqref{eq:eta}:
\begin{equation}
    \dot{N}_\mathrm{sig}=\rho_\mathrm{DM}VQ_L C\eta_{\rm trans}\begin{cases}
g_{a\gamma\gamma}^2 \frac{B_0^2}{m^2_a} ,&~\text{axion} ;\\
\frac13 \chi^2 , &~\text{dark photon}.
\end{cases} 
\end{equation}
By setting SNR equal to one we obtain 
\begin{equation}
\left(
\begin{array}{c}
    g_{a\gamma\gamma}^\mathrm{min}\\
    \chi^\mathrm{min}
\end{array}\right)
  = \sqrt{\frac{\mathrm{Max}\!\left[1,\,\dot{N}_\mathrm{noise}\,t\right]^{1/2}}
         {\rho_\mathrm{DM}\,V Q_LC\eta_\mathrm{trans}\,t}}
  \left(
\begin{array}{c}
     \frac{m_a}{B_0}\\
    \sqrt3
\end{array}\right)~.
\end{equation}

Since $V\propto\nu_{\rm DM}^{-3}$, in the noise-free or technical-noise-dominated regime, $g_{a\gamma\gamma}^{\rm min}(\chi^\mathrm{min})$ scales as $m_a^{5/2}(m_{A'}^{3/2})$. This is reflected in the linear part of each sensitivity curve for larger DM masses in Fig.~\ref{fig:sensitivity}. If the background is dominated by the thermal noise, which happens for smaller DM masses, the dependence of $g_{a\gamma\gamma}^{\rm min}(\chi^\mathrm{min})$ on $m_{\rm DM}$ reverses when $m_{\rm DM}$ exceeds $k_BT$. Each point on the sensitivity curves in Fig.~\ref{fig:sensitivity} corresponds to a separately optimized dielectric cavity matched to that target
frequency. Therefore, to cover $0.1$--$1$ THz would require $\mathcal{O}(10^5)$ discrete cavity configurations, given the cavity quality factor $Q_L\sim 10^4$. This is the principal practical cost of the resonant approach, as compared to a broadband approach.

% ==============================================================================
% Conclusion
% ==============================================================================
\section*{Conclusion and discussion}

We have proposed a hybrid detection architecture for ultralight dark matter in the $0.1$--$1.0$~THz range ($m_a \sim 0.4$--$4.1$~mev), combining a dielectric haloscope for resonant DM-to-THz conversion, Rydberg-atom six-wave mixing for coherent THz-to-optical transduction, and SNSPD readout. All three stages are integrated within a unified cryogenic platform at $\lesssim 1$~K. The directional and narrow-band nature of the atomic transducer helps suppress the isotropic thermal backgrounds. With 10 days of integration at $0.3$~K for each experiment, the setup shows the potential to reach the QCD axion band.

The sensitivity of the current scheme is primarily limited by the spectral mismatch between the cavity linewidth ($\Delta\nu_{\mathrm{cav}} \sim$~tens of MHz) and the atomic transition bandwidth ($\Delta\nu_{\mathrm{atomic}} \sim 1$~MHz), which reduces the bandwidth-averaged transduction efficiency to the percent level. Several avenues could substantially improve this. Higher cavity quality factors~\cite{Consolino_2018} would narrow $\Delta\nu_{\mathrm{cav}}$ and improve the spectral overlap, directly boosting $\eta_{\mathrm{trans}}$. Alternatively, power-broadening or multi-tone driving of the Rydberg transitions could widen $\Delta\nu_{\mathrm{atomic}}$ to better match the cavity line. Employing denser atomic ensembles could further enhance the internal conversion efficiency.

It is worth noting that the resonant and broadband Rydberg-based approaches to THz DM detection are complementary rather than competing. Broadband schemes such as BREAD with Rydberg readout~\cite{banerjee2025rydbergsinglephotondetection}  leverage large geometric collection areas and near-unity absorption efficiency to survey wide mass ranges simultaneously, while our architecture sacrifices broadband coverage for the coherent signal enhancement and directional thermal-noise rejection provided by the dielectric haloscope and SWM transducer. The resulting trade-off — scanning overhead versus coupling reach — makes broadband surveys efficient for initial exploration, while resonant searches are necessary to further probe the QCD axion band.

A natural experimental roadmap proceeds in two stages. The first targets dark photon dark matter, for which no external magnetic field is required. The second stage incorporates a high-field solenoid ($B_0 \sim 10$~T) for the full axion search, expanding to multiple target frequencies through interchangeable dielectric stacks and Rydberg transition schemes.
Each subsystem draws on rapidly maturing technologies: cryogenic atom trapping at sub-kelvin temperatures has been demonstrated with second-scale coherence times~\cite{kumar_quantum-enabled_2023,PRXQuantum.6.020337}, precision dielectric microstructure fabrication at THz scales is now routine~\cite{2017DielectricHaloscopesNewWayDetectAxionDarkMatter}, and SNSPDs with sub-$10^{-5}$~cps dark counts are commercially available~\cite{Zeuthen_2020}. The modular architecture further allows independent optimization of each stage, and even scaling to multi-cavity arrays for broader frequency coverage. This work establishes a concrete path toward exploring QCD axion parameter space in the meV mass range.

\subsection*{Acknowledgments}
C.G. is supported by NSFC under Grant No.~25201009.  
Y.Z. is supported by NSFC under Grant No.~92476103. J.F.C. acknowledges NSFC through Grants No.~92476102, and the Natural Science Foundation of Guangdong Province under Grant No.~2022B1515020096. 
J.S.'s work  is supported by NSFC through Grants No.~12450006 and No.~12025507. 
Y.M.Z. is supported by GRF Grants No.~11302824 and No.~11310925 from the Research Grants Council, University Grants Committee, and Grants No.~9610645 and No.~7020130 from the City University of Hong Kong. C.G. and Y.M.Z. acknowledge the Aspen Center for Physics, which is supported by NSF Grant No.~PHY-2210452 and partially supported by a grant from the Durand Fund, for their hospitality during the initialization of this study. 

% ==============================================================================
% Bibliography
% ==============================================================================

\bibliographystyle{apsrev4-2}

\bibliography{main.bib}

% ==============================================================================
% End Matter (PRL allows 1-2 pages after references)
% ==============================================================================

\setcounter{equation}{0}

\setcounter{figure}{0}
\renewcommand{\thefigure}{A\arabic{figure}}
\renewcommand{\theHfigure}{A\arabic{figure}}

\setcounter{table}{0}
\renewcommand{\thetable}{A\arabic{table}}
\renewcommand{\theHtable}{A\arabic{table}}

\appendix 
\onecolumngrid
\bigskip
\hrule
\bigskip
\section{Cavity mode design}\label{app:cavity}

We optimize the cavity geometry using 3D finite-element simulations, accounting for the loss tangent of silicon ($\tan\delta = 10^{-4}$) and copper surface impedance. A five-parameter scan (cavity base dimension, number of Si layers, coupling aperture radius, vacuum gap, and Si layer thickness) at 0.1~THz reveals optimal layer thicknesses of $d_\text{vac} = \nu^{-1}_\text{DM}/2$ and $d_\text{Si} = 1.12\,\nu^{-1}_\text{DM}/(2n_\text{Si})$ with $n_\text{Si} = 3.4$. These ratios are adopted as fixed scaling parameters for all target frequencies, reducing subsequent optimizations to three parameters. Figure~\ref{fig:efield} shows the resulting $E_x$ distribution at 0.1~THz, confirming strong coupling to a transverse $\bm{B}_0$ with $C =0.37$.

Table~\ref{tab:cavity_params} summarizes the optimized cavity geometry and
performance for benchmarks selected from $0.1$--$1$ THz frequencies. At each frequency, the cavity has a square base with side length 2$R_\mathrm{cav}$  and a height $L_\mathrm{cav}$. The loaded quality factor
$Q_L$ ranges from $1.7 \times 10^4$ at the lowest frequency to $7.5 \times
10^3$ at 1.0\,THz, reflecting increased dielectric and ohmic losses at shorter
wavelengths. The form factor $C \approx 0.37$--$0.40$ remains nearly
constant across the band, indicating that the fixed thickness ratios
($d_\mathrm{vac} = \nu^{-1}_\text{DM}/2$, $d_\mathrm{Si} = 1.12\,\nu^{-1}_\text{DM}/2n_\text{Si}$)
consistently produce band-edge modes with strong overlap to a uniform driving
field. The number of dielectric layers $N$ is 4 to 5 at most frequencies. For each target frequency, the table also lists the
nearest Rydberg transition ($|n_S\rangle \to |n_P\rangle$ in  $^{87}$Rb) and
the residual frequency offset $\delta f$; all offsets lie within $\pm
26$\,MHz, well within the $\sim$28\,MHz Zeeman tunability~\cite{2010QuantuminformationRydbergatoms}, confirming that
continuous cavity--atom frequency matching is achievable across the band.\\

\begin{table*}[h]
\caption{Optimized cavity parameters for selected benchmarks.
$N$ is the number of dielectric layers, $R_\mathrm{cav}$ the half length of each side of the square base, 
$L_\mathrm{cav}$ the cavity height, $Q_L$ the loaded quality factor,
$C$ the form factor for coupling to a transverse $\bm{B}_0$,
$\Delta\nu_\mathrm{cav}$ the cavity bandwidth, and $\delta f$ the frequency offset from the nearest Rydberg transition ($|n_S\rangle \to |n_P\rangle$).}
\label{tab:cavity_params}
\begin{ruledtabular}
\begin{tabular}{ccccccccc}
$f_\mathrm{target}$ [THz] & $N$ & $R_\mathrm{cav}$ [mm] & $L_\mathrm{cav}$ [mm] & $Q_L$ & $C$ & $\Delta\nu_\mathrm{cav}$ [MHz] & $\delta f$ [MHz] & Transition \\
\hline
0.107 & 5 & 2.82 & 9.32 & $1.73 \times 10^4$ & 0.37 & 6.2  & $+23.6$ & $34S \to 34P$ \\
0.200 & 5 & 1.50 & 4.98 & $1.38 \times 10^4$ & 0.37 & 14.5 & $+1.3$  & $57S \to 62P$ \\
0.298 & 4 & 1.03 & 2.67 & $1.20 \times 10^4$ & 0.38 & 24.8 & $-20.3$ & $58S \to 67P$ \\
0.400 & 5 & 0.75 & 2.49 & $1.05 \times 10^4$ & 0.37 & 38.2 & $-3.3$  & $40S \to 43P$ \\
0.501 & 4 & 0.61 & 1.59 & $9.72 \times 10^3$ & 0.38 & 51.5 & $-19.7$ & $54S \to 68P$ \\
0.602 & 4 & 0.52 & 1.32 & $9.16 \times 10^3$ & 0.38 & 65.7 & $-19.4$ & $32S \to 34P$ \\
0.700 & 4 & 0.45 & 1.14 & $8.59 \times 10^3$ & 0.38 & 81.6 & $-3.8$  & $48S \to 62P$ \\
0.805 & 4 & 0.39 & 0.99 & $8.08 \times 10^3$ & 0.38 & 99.7 & $-25.4$ & $32S \to 35P$ \\
0.901 & 4 & 0.34 & 0.88 & $7.52 \times 10^3$ & 0.38 & 119.8 & $+17.9$ & $33S \to 37P$ \\
1.003 & 3 & 0.32 & 0.60 & $7.50 \times 10^3$ & 0.40 & 133.8 & $-17.1$ & $35S \to 41P$ \\
\end{tabular}
\end{ruledtabular}
\end{table*}

\section{Rydberg Atom Transducer Conversion Efficiency Optimization}\label{app:transduction_eff}

To detect the weak terahertz signal induced by axion dark matter, we employ a coherent up-conversion process based on a six-level Rydberg atomic system. This mechanism acts as a transducer, converting a single THz photon ($\Omega_T$) into a near-infrared optical photon ($\Omega_L$) that can be detected by single-photon detectors.

The transduction is governed by a six-wave mixing (SWM) process. The atom is prepared in a ladder configuration where four strong auxiliary lasers ($\Omega_{A1}$ to $\Omega_{A4}$) bridge the energy gaps between the ground state $|1\rangle$, intermediate states, and high-lying Rydberg states, namely $|1\rangle \to |2\rangle$, $|2\rangle \to |3\rangle$, $|4\rangle \to |5\rangle$, and $|5\rangle \to |6\rangle$. The input THz field $\Omega_T$ drives the transition $|3\rangle \rightarrow |4\rangle$, while the resulting optical signal $\Omega_L$ is emitted on the $|6\rangle \rightarrow |1\rangle$ closure transition. 
Under the rotating-wave approximation, the Hamiltonian in the interaction picture is:

\begin{equation}
\begin{split}
    H =& -\Delta_{A1} |2\rangle\langle2| - (\Delta_{A1} + \Delta_{A2}) |3\rangle\langle3| 
    - (\Delta_{A1} + \Delta_{A2} + \Delta_T) |4\rangle\langle4| \\
    &- (\Delta_{A1} + \Delta_{A2} + \Delta_T + \Delta_{A3}) |5\rangle\langle5| 
    - (\Delta_{A1} + \Delta_{A2} + \Delta_T + \Delta_{A3} - \Delta_{A4}) |6\rangle\langle6| \\
    &- \frac{1}{2} ( \Omega_{A1} |2\rangle\langle1| + \Omega_{A2} |3\rangle\langle2| + \Omega_{T} |4\rangle\langle3| 
    + \Omega_{A3} |5\rangle\langle4| + \Omega_{A4} |5\rangle\langle6| + \Omega_L |6\rangle \langle 1| + \text{H.c.} ),
\end{split}
\end{equation}
where $\Delta_X = \omega_X - \omega_{ij}$ is the frequency detuning of the laser frequency driving $\ket{i}\to\ket{j}$ transition. Rabi frequencies $\Omega_X$ represent the coupling strength between the laser light and the atomic transitions. Rabi frequencies are defined as $\Omega_X = \varepsilon_X d_{ij}/\hbar$, where $\varepsilon_X$ is the electric field amplitude, $d_{ij}$ is the transition dipole moment between states $|i\rangle$ and $|j\rangle$, and $\hbar$ is the reduced Planck constant. %The decay processes are modeled via a Lindblad master equation:
The complete physical picture of the atomic evolution, including both coherent driving and incoherent decay, is described by the Lindblad master equation:
\begin{equation}
    \dot{\rho}(z,t) = -\frac{i}{\hbar}[H, \rho] + \sum_{k} \left( \mathcal{L}_k \rho \mathcal{L}_k^\dagger - \frac{1}{2} \{\mathcal{L}_k^\dagger \mathcal{L}_k, \rho\} \right),
\end{equation}
with Lindblad operators $\mathcal{L}_{\Gamma,k} = \sqrt{\gamma_{k+1}}\ket{k}\bra{k+1}$ for spontaneous emission and $\mathcal{L}_{D,k'} = \sqrt{\gamma_{k'}}\ket{k'}\bra{k'}$ for dephasing. The latter accounts for the loss of coherence due to laser linewidth, atomic collisions, or environmental fluctuations. The steady-state solution is obtained by setting $\dot{\rho} = 0$. %The field amplitudes satisfy:

As the THz signal (axion-induced) and auxiliary beams propagate through the atomic cloud of length $L$, the transduction of the optical signal is described by the propagation equations:

\begin{equation}
\frac{\partial\Omega_{T}(z,t)}{\partial z} = i\kappa_{T}\rho_{43}(z,t), \quad
\frac{\partial\Omega_{L}(z,t)}{\partial z} = i\kappa_{L}\rho_{61}(z,t)\,.
\end{equation}
Here, we define $\kappa_{T(L)} = n_{at}|d_{ji}|^{2}\omega_{T(L)}/(\hbar\epsilon_{0}c)$ as the propagation coupling coefficient, where $n_{at}$ is the atomic density.

The internal conversion efficiency is defined as the ratio of emitted photon flux to input photon flux: $\eta_\text{int}(\omega_T) = \frac{|\Omega_L(z = L)|^2/\kappa_L}{|\Omega_T(z = 0)|^2/\kappa_T}$. For the benchmark study, we adopt the implementation based on the $6S_{1/2}$, $6P_{1/2}$, and $6P_{3/2}$ lower manifold of $^{87}$Rb; the optimized control parameters for each benchmark transition are listed in Table~\ref{tab:control_params}.
In the steady-state limit ($\dot{\rho}=0$) and under weak-signal conditions, the system can be optimized to achieve $ \eta_\text{int}^{\max}\gtrsim 85\%$ as shown in Table~\ref{tab:atom_params}. 

The bandwidth-averaged efficiency is approximately $\langle \eta_\text{int}\rangle\approx\eta_\text{int}^{\max}  \cdot (\Delta \nu_{\mathrm{atomic}}/\Delta \nu_{\mathrm{cav}})$, while the benchmark values in Table~\ref{tab:atom_params} are obtained from the full numerical scan.

\begin{table*}[t]
\caption{Optimized control parameters for the six-wave-mixing transduction scheme at different target THz frequencies. $f_{\mathrm{target}}$ is the target THz frequency corresponding to the transition ($|n_S\rangle \to |n_P\rangle$), $\Omega_{A1}$--$\Omega_{A4}$ are the Rabi frequencies of the four auxiliary driving fields, and $(\Delta_{A1},\Delta_{A2})$ are the frequency detunings of the first two auxiliary fields.}
\label{tab:control_params}
\begin{ruledtabular}
\resizebox{\textwidth}{!}{%
\begin{tabular}{ccccccc}
$f_{\mathrm{target}}$ [THz] &
Transition &
$\Omega_{A1}/2\pi$ [MHz] &
$\Omega_{A2}/2\pi$ [MHz] &
$\Omega_{A3}/2\pi$ [MHz] &
$\Omega_{A4}/2\pi$ [MHz] &
$(\Delta_{A1},\Delta_{A2})/2\pi$ [MHz] \\
\hline
0.107 & $34S \to 34P$ & 1.874  & 20.283 & 3.800 & 15.445 & $(0.691,-2.362)$ \\
0.200 & $57S \to 62P$ & 7.682  & 9.509  & 3.565 & 6.473  & $(-3.393,2.904)$ \\
0.298 & $58S \to 67P$ & 9.744  & 9.250  & 3.874 & 5.736  & $(-1.697,1.452)$ \\
0.400 & $40S \to 43P$ & 9.624  & 16.847 & 5.158 & 11.517 & $(-3.393,2.904)$ \\
0.501 & $54S \to 68P$ & 12.444 & 3.906  & 5.400 & 5.605  & $(-3.056,3.051)$ \\
0.602 & $32S \to 34P$ & 5.472  & 22.454 & 2.950 & 15.445 & $(-5.327,4.254)$ \\
0.700 & $48S \to 62P$ & 12.210 & 4.435  & 5.341 & 6.473  & $(-3.707,3.669)$ \\
0.805 & $32S \to 35P$ & 13.143 & 24.407 & 5.149 & 16.019 & $(-4.819,4.387)$ \\
0.901 & $33S \to 37P$ & 11.405 & 7.253  & 3.749 & 9.137  & $(-4.519,4.387)$ \\
1.003 & $35S \to 41P$ & 11.405 & 7.555  & 3.749 & 8.786  & $(-4.819,4.387)$ \\
\end{tabular}%
}
\end{ruledtabular}
\end{table*}

\subsection{Estimation of Spatial Mode-Matching Efficiency}\label{app:F}

The phase-matching envelope factor $F$ characterizes the coherent summation of signals generated across the finite volume of the Rydberg ensemble. Unlike the plane-wave approximation, the focused THz beam introduces intrinsic phase variations—most notably the Gouy phase shift and wavefront curvature—that can lead to destructive interference even under perfect wave-vector matching ($\Delta k = 0$). 
To quantify this, we adopt the Bjorklund formalism for nonlinear interactions of focused Gaussian beams~\cite{nonlinearoptics}. Given that only THz signal is focused, $F$ is defined as the normalized magnitude of the complex spatial overlap integral:
\begin{equation}
\label{eq:F}
    F(\zeta) = \frac{|J(\zeta)|^2}{\zeta^2},
\end{equation}
where $\zeta = L/z_R$ is the focusing parameter (the ratio of the ensemble thickness $L$ to the THz Rayleigh range $z_R$). The integral $J(\zeta)$ is given by:

\begin{equation}
    J(\zeta) = \int_{-\zeta/2}^{\zeta/2} \frac{1}{(1+i\tau)^2}  \mathrm{d}\tau~.
\end{equation}
The term $(1+i\tau)^{-2}$ represents the longitudinal amplitude and phase evolution of the Gaussian mode. As $\zeta$ increases (i.e., as the THz Rayleigh range $z_R$ becomes shorter relative to $L$), the non-linear phase slip (Gouy phase) between the driving polarization and the generated signal field increases, leading to the reduction of $F$.

\begin{figure}[h]
    \centering
    \includegraphics[width=0.5\linewidth]{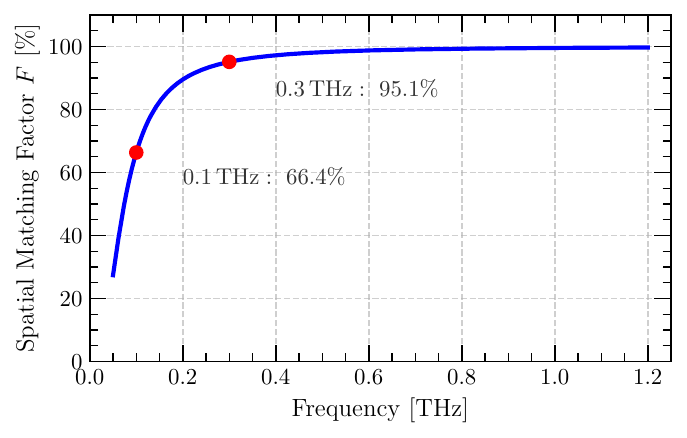}
    \caption{$F$ (Eq.~\ref{eq:F}) as a function of THz frequency.}
    \label{fig:spatial}
\end{figure}

For our experimental parameters ($w_0 = 1.0$ mm, $L = 1.0$ mm), the THz Rayleigh range $z_R = \pi w_0^2 / \lambda_T$ scales linearly with frequency. This causes the spatial factor $F$ to exhibit a strong frequency dependence, as Fig.~\ref{fig:spatial} shows, particularly for the low frequencies.

\begin{table*}[h]
\caption{Benchmark transduction performance for different target frequencies. $f_{\mathrm{target}}$ is the target THz frequency corresponding to the transition ($|n_S\rangle \to |n_P\rangle$), $\Delta\nu_{\mathrm{cav}}$ is the cavity bandwidth, $\Delta\nu_{\mathrm{atomic}}$ is the atomic transition linewidth, $F$ is the phase-matching envelope factor, $\eta_{\rm int}^{\rm max}$ is the maximum internal conversion efficiency, $\langle\eta\rangle$ is the bandwidth-averaged internal efficiency, and $\eta_{\rm trans}$ is the total transduction efficiency.}
\label{tab:atom_params}
\begin{ruledtabular}
\begin{tabular}{cccccccc}
$f_\mathrm{target}$ [THz] & $\Delta\nu_\mathrm{cav}$ [MHz] & Transition & $\Delta\nu_\mathrm{atomic}$ [MHz] & $F$ & $\eta_{\rm int}^{\rm max}$ [\%] & $\langle\eta\rangle$ [\%] & $\eta_{\rm trans}$ [\%] \\
\hline
0.107 & 6.17   & $34S \to 34P$ & 4.067 & 0.6635 & 89.8563 & 26.7754 & 17.7658 \\
0.200 & 14.53  & $57S \to 62P$ & 1.910 & 0.8952 & 94.0872 & 6.5282  & 5.8441  \\
0.298 & 24.77  & $58S \to 67P$ & 1.185 & 0.9513 & 94.1532 & 2.4457  & 2.3265  \\
0.400 & 38.22  & $40S \to 43P$ & 3.263 & 0.9721 & 96.4511 & 4.6399  & 4.5106  \\
0.501 & 51.55  & $54S \to 68P$ & 1.906 & 0.9820 & 86.2394 & 1.7473  & 1.7159  \\
0.602 & 65.68  & $32S \to 34P$ & 2.306 & 0.9875 & 97.2476 & 2.1034  & 2.0770  \\
0.700 & 81.58  & $48S \to 62P$ & 1.858 & 0.9908 & 85.1554 & 1.0633  & 1.0535  \\
0.805 & 99.73  & $32S \to 35P$ & 3.626 & 0.9929 & 97.1859 & 2.0643  & 2.0497  \\
0.901 & 119.78 & $33S \to 37P$ & 3.590 & 0.9944 & 88.2923 & 1.5864  & 1.5775  \\
1.003 & 133.78 & $35S \to 41P$ & 2.789 & 0.9955 & 87.4532 & 1.0360  & 1.0313  \\
\end{tabular}
\end{ruledtabular}
\end{table*}

\section{Thermal noise suppression via directional transduction}\label{app:xi} 

The key advantage of Rydberg-atom transduction for DM detection is its
intrinsic rejection of isotropic thermal backgrounds through phase matching.
We quantify this suppression following the framework of Ref.~\cite{borowka_continuous_2024}.

A thermal field at temperature $T$ and frequency $\omega$ has an isotropic
spectral energy density with photon occupation number $\bar{n} = (e^{\hbar\omega/k_BT}-1)^{-1}$. The effective thermal field amplitude seen by the atomic transducer is not the full blackbody field, but
is filtered by the angular response of the SWM process:
\begin{equation}
  \langle E_\mathrm{eff}^2 \rangle
  = \frac{2\omega^2 \hbar\omega\,\bar{n}}{\pi^2 c^3 \varepsilon_0}
  \cdot \frac{1}{4\pi} \int d\Omega\, \frac{|\eta_\mathrm{noise}(\theta)|^2}2\,,
  \label{eq:Eeff}
\end{equation}
where the angular response function
\begin{equation}
  |\eta_\mathrm{noise}(\theta)|^2
  = \left[\cos^4(\theta/2) + \sin^4(\theta/2)\right]
  \cdot \mathrm{sinc}^2\left(\Delta k(\theta) L/2\right)\,,~ \Delta k(\theta) = [1-\cos{(\theta)}]\omega_T /c 
\end{equation}
encodes two distinct filtering mechanisms. The first factor captures the
polarization-dependent coupling of the thermal field to the atomic dipole
transition. The second factor $\mathrm{sinc}^2(\Delta k\, L/2)$ enforces phase-matching selectivity. The phase mismatch factor, $\Delta k$, quantifies the vector sum deviation among the six wave vectors involved in the SWM process. Given that the five driving fields are aligned in a strict collinear configuration, the variation of $\Delta k$ is primarily determined by the orientation of the thermal noise. Since this thermal noise is intrinsically isotropic, it necessitates an angular integration over the noise wave vectors to accurately determine the total generated signal intensity. Here the normalization is chosen such that $\frac12 \langle E_\mathrm{eff}^2 \rangle$ gives the energy density per angular frequency.

The resulting thermal photon rate at the detector is thus given by
\begin{equation}
  \dot{N}_\mathrm{thermal}
  = \frac{\langle E_\mathrm{eff}^2 \rangle\,
          \Delta\nu_\mathrm{atomic}\,\pi \varepsilon_0 c\, A_\mathrm{eff}}
         {\hbar\omega_T}\,,
  \label{eq:Nthermal}
\end{equation}
where $\Delta \nu_\mathrm{atomic} \approx 1\,\text{MHz}$ is the atomic
detection bandwidth and $A_{\rm eff}\approx\pi (0.5\,\mathrm{mm})^2$ is the effective transduction area. Compared to a hypothetical
omnidirectional detector of the same bandwidth, the directional SWM transducer suppresses the thermal rate by a factor $\xi(\omega L)\equiv \frac{1}{4\pi} \int d\Omega\, \frac{|\eta_\mathrm{noise}(\theta)|^2}2$, which is given by
\begin{equation}
  \xi (\alpha)
  =\frac{-2 \gamma  \alpha +\alpha -\sin (2 \alpha )+\alpha  \cos (2 \alpha )+2 \alpha  (-\log (2 \alpha )+\text{Ci}(2 \alpha )+\alpha  \text{Si}(2 \alpha ))}{4 \alpha ^3} \,,
  \label{eq:xi}
\end{equation}
where $\text{Si}$ and $\text{Ci}$ are sine integral and cosine integral functions, respectively.

\begin{figure}
    \centering
    \includegraphics[width=0.5\linewidth]{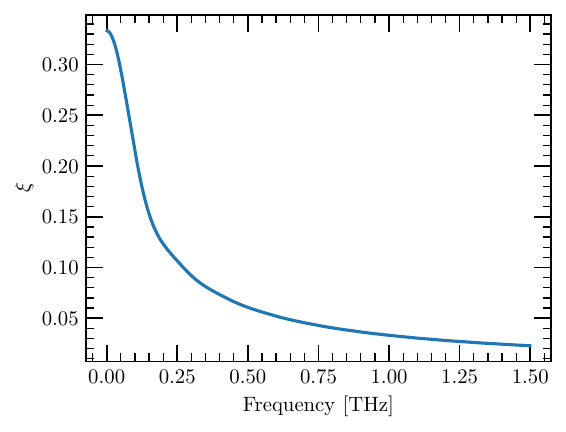}
    \caption{Suppression factor $\xi$ \eqref{eq:xi} as a function of thermal noise frequency. $L=1$~mm.}
    \label{fig:suppression}
\end{figure}

\begin{figure}
    \centering
    \includegraphics[width=0.5\linewidth]{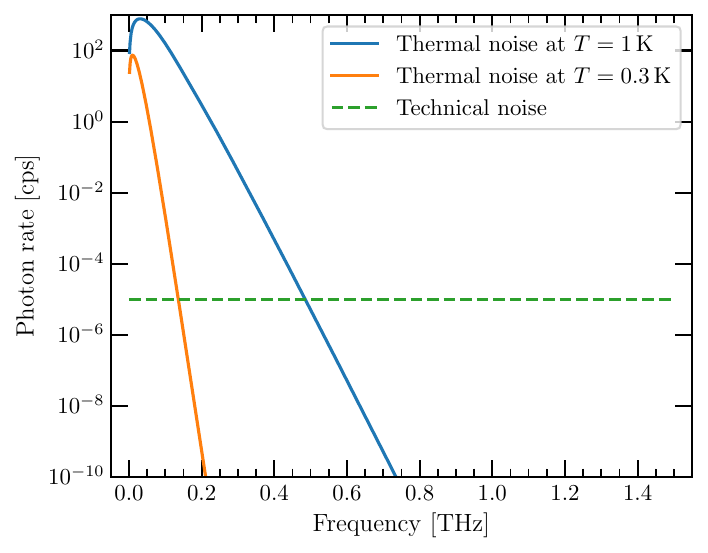}
    \caption{Thermal noise ($T=1 \,\text{K}$ and 0.3 K) and technical noise versus frequency.}
    \label{fig:noise}
\end{figure}

Fig.~\ref{fig:noise} shows the dominant noise sources, including thermal noise and total technical noise.
Above frequencies corresponding to 1 K, the noise budget is still dominated by thermal photons.

\end{document}